\documentclass[a4paper,11pt]{article}
\usepackage{amsmath,amssymb,amsthm}
\usepackage{amsfonts}
\usepackage{eucal}

\renewcommand{\d}{\delta}

\renewcommand{\l}{\lambda}

\renewcommand{\L}{\Lambda}
\newcommand{\co}{\mathbb{C}}
\newcommand{\re}{\mathbb{R}}
\newcommand{\ze}{\mathbb{Z}}

\newcommand{\HKSh}{H_{\mathrm{KS};h}}
\newcommand{\HHDh}{H_{\mathrm{HD};h}}
\newcommand{\HHDtwoh}{H_{\mathrm{HD};2h}}

\newcommand{\bigset}[2]{\bigl\{#1\bigm|#2\bigr\}}

\newcommand{\norm}[1]{\| #1 \|}

\newtheorem{thm}{Theorem}

\theoremstyle{definition}

\theoremstyle{remark}

\title{Equivalence of the staggered fermion Hamiltonan\\ and the discrete Hodge-Dirac operator\\ on square lattices}
\author{Shu Nakamura\footnote{Department of Mathematics, Faculty of Sciences, Gakushuin University, 1-5-1, Mejiro, Toshima, Tokyo, Japan 171-8588, \texttt{shu.nakamura@gakushuin.ac.jp}. 
The research is partly supported by JSPS Kakenhi Grant, Kiban (C) 21K03276} }

\begin{document}
\maketitle

\begin{abstract}
We show that the free massless staggered fermion (or the KS-fermion) Hamiltonian is equivalent to a 
discrete Hodge-Dirac operator on the $d$-dimensional square lattice $h\ze^d$. In fact, they are identical  
operator valued matrices 
under suitable choices of their representations on $\ell^2(2h\ze^d)\otimes\co^{2^d}$. We employ the formulations  
of the staggered fermion by Nakamura (2024), and the discrete cohomology structure on the square lattices 
by Miranda-Parra (2023). 
\end{abstract}

\section{Introduction}\label{sec-intro}

In this short note, we show that the massless staggered fermion Hamiltonian (Susskind~\cite{Susskind}, 
see also \cite{KS} and \cite{Rothe}) is essentially the same operator 
as a discrete analogue of the Hodge-Dirac operator on the Euclidean space. 

The staggered fermion model (or the KS-fermion model) is widely used in the lattice field 
theory to describe the Dirac particle on the lattice, possibly with additional symmetry (see \cite{Rothe}). 
A mathematical general formulation and its continuum limit are discussed by Nakmaura~\cite{Na}. 
On the other hand, the Hodge-Dirac operator $d+d^*$ on manifolds is studied comprehensively in differential geometry, 
and a discrete analogue is discussed by Miranda and Parra~\cite{Miranda-Parra}. 
They both are represented on the $2^d$-dimensional vector space valued functions on $\ze^d$, 
and they have similar properties, in particular, their squares are the Laplace operators. 
Thus it is natural to expect that they have close relationship to each other, and we confirm it in this note. 

We set $d\geq 1$ be the space dimension. 
We denote $h>0$ be the mesh size of our lattice, and we consider function spaces on 
either $h\ze^d$ or $2h\ze^d$. 
Let $\HKSh$ be the massless KS-Hamiltonian defined on $(\ell^2(2h\ze^d))^{2^d}$, 
as in \cite{Na} Subsection~5.1. 
We note $\HKSh$ is a bounded self-adjoint operator. 
We describe this operator in more detail in the next section. 

Let $X(h\ze^d)=\bigcup_{j=0}^d X^j(h\ze^d)$ be the set of the combinatorial differential complex
on $h\ze^d$ as defined in Section~4 of \cite{Miranda-Parra}, and let $d$ 
be the discrete exterior derivative on $\ell^2(X)$. We denote 
\[
H_{\mathrm{HD};h} =-i(d-d^*)
\]
be our Hodge-Dirac operator on $\ell^2(X)$. We note $\HHDh$ is slightly different from the standard 
Hodge-Dirac operator: $d+d^*$, though they are unitarily equivalent. 
We will also describe them in more detail in the next section. 


\begin{thm}\label{main-thm}
There is a unitary transform $U$ : $\bigoplus_{k=0}^d\bigwedge^k \co^d\simeq \co^{2^d} \to \co^{2^d}$ such that 
\[
UH_{\mathrm{HD};2h} U^* = \HKSh. 
\]
\end{thm}

For the proof, we construct suitable orthonormal basis for the both systems so that the 
operator-valued matrix representations of the operators $\HKSh$ and $H_{\mathrm{HD};2h}$ are identical. 

The discrete Dirac operators has been extensively used in the lattice field theory (see, e.g., \cite{Rothe}), 
but they have been not necessarily well-known in mathematics, or even in mathematical physics. 
Recently, in relation to the continuum limit (\cite{NaTad}), the discrete Dirac operators have attracted some attention 
(\cite{C-G-J-2}, \cite{Na}, \cite{Schmidt-Umeda}) 
and the Hodge-Dirac operator was also studied (\cite{Miranda-Parra}) in relation to the continuum limit. 
Recently, discrete Dirac operators has also been studied in relation to the network theory (see, e.g., \cite{Bianconi24}
and references therein). We also note that the discrete versions of the Nambu--Jana-Lasinio model 
employ the staggered fermion effectively, and the mass generation of these models is actively studied 
(see Goto-Koma \cite{Goto-Koma-23} and also \cite{Bianconi24}). 

\section{Proof}
\subsection{Notations}

We collect several notations for function spaces, operators, which are used in the following 
discussion. We denote the square lattice with the mesh $h>0$ by 
\[
h\ze^d =\bigset{hn}{n\in\ze^d}
\]
and the standard basis of $\re^d$ by $\{e_1,\dots,e_d\}$, i.e., $(e_j)_k=\d_{j,k}$, 
$k=1,\dots,d$,  where $\d_{j,k}$ is the Kronecker symbol. 

Our function spaces are mostly square summable function spaces $\ell^2(h\ze^d)$, and 
we suppose $\ell^2(h\ze^d)$ is equipped with the norm
\[
\norm{u}_{\ell^2(h\ze^d)}^2 = h^d\sum_{z\in h\ze^d} |u(z)|^2, \quad u\in\ell^2(h\ze^d).
\]
We note that we often replace $h$ by $2h$, and the norm changes accordingly. 

For a function $u$ on the lattice $h\ze^d$, we apply the following difference operators. 
The symmetric difference operators are defined by 
\[
D_{h;j}^S u(z) =\frac{1}{2ih}(u(z+he_j)-u(z-he_j)), 
\]
where $z\in h\ze^d$ and $j=1,\dots, d$. 
The forward and backward difference operators are 
\[
D_{h;j}^\pm u(z)= \pm\frac{1}{ih}(u(z\pm he_j)-u(z)).  
\]
We note
\begin{equation}\label{eq:key-diff-formula}
D_{h;j}^S u(z+ h e_j)= D_{2h;j}^+ u(z), \quad
D_{h;j}^S u(z)= D_{2h;j}^- u(z+ he_j),
\end{equation}
and this observation plays an essential role in the representation of the KS fermion model. 

\subsection{The staggered fermion model}

Here we recall the construction of the operator $\HKSh$ following \cite{Na}, 
and modify it so that our theorem becomes obvious. 

We write
\[
s_j(n)=\sum_{k=1}^j n_k \quad \text{for }n\in\ze^d, j=1,\dots,d, 
\]
and set $s_0(n)=0$. Then the massless KS-fermion Hamiltonian is originally defined by 
\[
\tilde H_{\mathrm{KS};h} u(z) =\sum_{j=1}^d (-1)^{s_{j-1}(z/h)} D_{h;j}^S u(z), \quad z\in h\ze^d, 
\]
for $u\in\ell^2(h\ze^d)$. We decompose $\ell^2(h\ze^d)$ into a $2^d$ direct sum of $\ell^2(2h\ze^d)$, 
and we obtain an operator $\HKSh$ on $[\ell^2(2h\ze^d)]^{2^d}$ (without the 
fermion doubling problem, see \cite{Na}). Let 
\[
\L =\{0,1\}^d=\bigset{a\in\ze^d}{a_k=0, 1, k=1,\dots, d}
\]
be the index set so that each point in $z\in h\ze^d$ is represented as $z=w+h a$ with $w\in 2h\ze^d$
and $a\in\L$. We set a unitary operator $U_h$ : 
$\ell^2(h\ze^d)\to [\ell^2(2h\ze^d)]^{\L}=\ell^2(2h\ze^d)\otimes\co^\L$ by 
\[
(U_h u)_a(z)= 2^{-d/2} u(z+ha), \quad z\in 2h\ze^d, a\in \L,
\]
and we define the KS-fermion Hamiltonian by
\[
\HKSh =U_h \tilde H_{\mathrm{KS};h} U_h^* \quad \text{on }\ell^2(2h\ze^d)\otimes\co^{\L}.
\]
We recall the matrix elements of $\HKSh$ is given explicitly by 
\[
(\HKSh)_{a,b} =\begin{cases} (-1)^{s_{j-1}(a)}D_{2h;j}^+, \quad &\text{if }b=a-e_j,\\
(-1)^{s_{j-1}(a)}D_{2h;j}^- ,\quad &\text{if }b=a+e_j,\\ 
0, \quad&\text{otherwise}., \end{cases}
\]
thanks to the formula \eqref{eq:key-diff-formula}. 
We note $s_{j-1}(a\pm e_j)=s_{j-1}(a)$, and hence we may replace $s_{j-1}(a)$ 
by $s_{j-1}(b)$ in the above expression of $\HKSh$. Then it is easy to see that this operator is symmetric. 

We refine the above decomposition as follows. 
For $a\in\L$, we write $|a|=a_1+\cdots+a_d$, and 
\[
\L=\bigcup_{k=0}^d \L_k, \quad \L_k=\bigset{a\in\L}{|a|=k}, \quad k=0,\dots, d. 
\]
For $a\in \L_k$, we may consider $z=w+ha$ represents a $k$-simplex in $2h\ze^d$: 
\[
\l_a(w)=\prod_{j=1}^d [w_j,w_j+2h a_j]
\]
with the positive orientation. Thus each element of $2h\ze^d\otimes\L_k$ represents 
a $k$-simplex over $2h\ze^d$. Now we may consider $\pmb{C}^k(2h\ze^d)=\ell^2(2h\ze^d)\otimes{\co^{\L_k}}$ 
as the space of (square summable) $k$-cochain over $2h\ze^d$, and 
\[
\pmb{C}(2h\ze^d)=\ell^2(2h\ze^d)\otimes{\co^\L}= \bigoplus_{k=0}^d \pmb{C}^k(2h\ze^d)
\]
is the space of the square summable cochains. 

We then set an operator $\pmb{d}$ on $\ell^2(2h\ze^d)\otimes{\co^\L}$ by the matrix elements as follows. 
\[
(\pmb{d})_{a,b} =\begin{cases}  (-1)^{s_{j-1}(a)}iD_{2h;j}^+, \quad &\text{if }b=a-e_j,\\
0, \quad&\text{otherwise}. \end{cases}
\]
It is easy to observe that $\pmb{d}$ maps $\pmb{C}^j(2h\ze^d)$ to $\pmb{C}^{j+1}(2h\ze^d)$, 
$j=0,\dots, d-1$, 
and it has properties similar to the exterior derivative. Moreover, it is also easy to see 
\[
\HKSh =(-i\pmb{d})+(-i\pmb{d})^*=-i(\pmb{d}-\pmb{d}^*).
\]
We will see that the discrete Hodge-Dirac operator on $2h\ze^d$ has the same representation. 

\subsection{The discrete Hodge-Dirac operators}

We use essentially the same notation as in Miranda-Parra~\cite{Miranda-Parra}, though we do not use 
the simplexes with negative orientations. 
We note an element of $\pmb{C}^0(2h\ze^d)$ is simply a square summable function on $2h\ze^d$. 
For $\pmb{C}^1(2h\ze^d)$, each element can be written as 
\[
f^1=\sum_{j=1}^d f_j^1(z)dx^j(z),
\]
where $f_j^1\in\ell^2(2h\ze^d)$, and $\{dx_j(z)\mid z\in2h\ze^d,j=1,\dots,d\}$ is the dual basis 
to the basis $\{\l_j(z)=[z,z+2he_j]\mid z\in2h\ze^d,j=1,\dots,d\}$ of the 1-chain. 
For general $a\in\L_k$, we denote 
\[
dx^a=dx^{j_1}\wedge dx^{j_2}\wedge\cdots\wedge dx^{j_k}
\]
where $\{j_1,\dots, j_k\}=\{j\,|\,a_j=1\}$ and $j_1<j_2<\dots<j_k$. 
We note $\{dx^a(z)\}_{a,z}$ is the dual basis of $\{\l_a(z)\}_{a,z}$, and 
each element in $\pmb{C}(2h\ze^d)=\ell^2(2h\ze^d;\bigwedge\co^d)$ is represented as 
\[
f=\sum_{a\in\L} f^a(z) dx^a(z),\quad f^a\in \ell^2(2h\ze^d). 
\]
We compute the exterior derivative 
using the formula (20) of \cite{Miranda-Parra}, which is a natural analogue of the definition of 
the exterior derivative for smooth manifolds. Let $b\in\L_k$, then we have
\[
\tilde d_k (f^b dx^b)=(\tilde d_0 f^b)\wedge dx^b 
=\sum_{j=1}^d \frac{1}{2h}(f^b(z+2he_j)-f^b(z))dx^j\wedge dx^b.
\]
Now we note $dx^j\wedge dx^b\neq 0$ if and only if $a=b+e_j\in\L$, and then 
\[
dx^j\wedge dx^b= (-1)^{s_{j-1}(b)}dx^a
\]
since $s_{j-1}(b)=s_{j-1}(a)$ counts the number of non-zero entry of $b$ less than $j$. 
Hence we can write 
\[
\tilde d_k(f^bdx^b) =\sum_{a=b+e_j\in\L} (-1)^{s_{j-1}(a)}(iD_{2h;j}^+ f^b)dx^{a}.
\] 
Thus $\tilde d$ has the exactly same expression as $\pmb{d}$ with respect to 
the corresponding orthonormal basis. In other words, as operator valued $2^d\times 2^d$ 
matrices, $\HHDtwoh$ and $\HKSh$ are exactly the same operator, and in particular, 
Theorem~\ref{main-thm} follows. \qed


\end{document}